# The transmission spectrum of Earth through lunar eclipse observations

Enric Pallé[1], María Rosa Zapatero Osorio[1], Pilar Montañés-Rodríguez[1], Rafael Barrena[1], Eduardo L. Martín[1,2]

[1] Instituto de Astrofísica de Canarias, Vía Láctea s/n, E38205 La Laguna, Tenerife, Spain.

[2] University of Central Florida, Physics Department, P.O. Box 162385, Orlando, FL32816, USA

Of the 342 planets discovered[1] so far orbiting other stars, 58 'transit' the stellar disk, meaning that they can be detected by a periodic decrease in the starlight flux[2]. The light from the star passes through the atmosphere of the planet, and in a few cases the basic atmospheric composition of the planet can be estimated[3-5]. As we get closer to finding analogues of Earth[6-8], an important consideration toward the characterization of exoplanetary atmospheres is what the transmission spectrum of our planet looks like. Here we report the optical and near-infrared transmission spectrum of the Earth, obtained during a lunar eclipse. Some biologically relevant atmospheric features that are weak in the reflected spectrum[9] (such as ozone, molecular oxygen, water, carbon dioxide and methane) are much stronger in the transmission spectrum, and indeed stronger than predicted by modelling[10-11]. We also find the fingerprints of the Earth's ionosphere and of the major atmospheric constituent, diatomic nitrogen ($N_2$), which are missing in the reflected spectrum.

The characterization of spectral features in our planet's transmission spectrum can be achieved through observations of the light reflected from the Moon toward the Earth during a lunar eclipse, which resembles the observing geometry during a planetary transit. At that time, the reflected sunlight from the lunar surface within the Earth's umbra will be entirely dominated by the fraction of sunlight that is transmitted through an atmospheric ring located along the Earth's day-night terminator (see Figure S1). Observations of the lunar eclipse in August 16[th] 2008 have allowed us to characterize the Earth's spectrum as if it were observed from an astronomical distance during a transit in front of the Sun. Except for some early attempts[12-13] with photographic plates, spectroscopic lunar eclipse observations in the visible or the near-infrared wavelengths, have never been undertaken.

The Earth transmission spectrum can be calculated from the brightness ratio of the light reflected by the lunar surface when in the umbra, penumbra and out of the eclipse (see the on-line material). The resulting transmission spectrum is shown in Figure 1a, where simultaneous optical and near-infrared observations with the William Herschel and Nordic Optical Telescopes are inter-calibrated to provide continuous wavelength coverage from 0.36 to 2.40 ⬜m. It is common knowledge that the Earth optical and near-infrared transmission spectrum is red[14], i.e., more solar flux successfully passes through the atmosphere at longer wavelengths, as can be inferred form simple naked-eye observations of a lunar eclipse, or of a sunset/sunrise. The rising nature of the transmission spectrum continuum (possibly the most remarkable feature of the data shown in Figure 1a) is caused by the Rayleigh scattering of air, which in addition to the ozone Chappuis band absorption between 0.375 and 0.650 ⬜m is rather efficient in scavenging short wavelength radiation through a long atmospheric path.

During a transit, the starlight travels through a much larger airmass of the planet's atmosphere than in reflection. This causes the two spectra to be different (see Figure 1b). In transmission, the Earth is brighter in the near-infrared range, particularly at 2.2 μm. In contrast, the Earth's reflected spectrum appears blue[15] because the very same Rayleigh scattering effect expels out the short wavelength radiation back to space. Thus, in transmission, the pale blue dot becomes the pale red dot.

Moreover, the transmission spectrum presents strong absorption features produced by molecular oxygen and oxygen collision complexes, including collisions between $O_2$ and $N_2$, which are significantly more intense than in the reflectance spectrum. Oxygen collision complexes are van der Waals molecules, also know as dimers[16-18], which can be used in combination with other molecular oxygen bands to derive an averaged atmospheric column density of $N_2$. This is important because although $N_2$ is the major atmospheric component (78.08 % in volume), it lacks any marked electronic transition. The strength of these bands in the transmission spectrum implies that atmospheric dimers may become a major subject of study for the interpretation of rocky exoplanets transmission spectra and their atmospheric characterization. The atmospheric spectral bands of $O_3$, $O_2$, $H_2O$, $CO_2$, and $CH_4$ are readily distinguishable in the transmission spectrum (see Figure S3). Trace amounts of $N_2O$, $OClO$ and $NO_2$ (a gas mainly produced by human activities) might also be present in our data, but their detection will need future detailed modelling efforts to fit the observations.

The presence of the Earth's ionosphere is also revealed in our transmission spectrum through the detection of relatively weak and narrow absorption lines corresponding to the singly ionized calcium atoms (Ca II, see Figure S3). Calcium is the sixth most abundant element on Earth.. It is possible that other, more abundant[19-20] ionospheric species, such as the singly ionized magnesium (Mg II), can be detectable at shorter wavelengths that are not covered by our data. The neutral atomic resonance doublet of sodium (Na I) is embedded in a quite strong and broad absorption due to dimers, ozone and Rayleigh scattering, and only the core of the doublet is detected at 0.5898 ᴗm.

In the coming years, space missions such as the James Webb Space Telescope, will perhaps have the opportunity to perform transit spectroscopy of rocky planets[21,22]. The faint signal of the planet atmosphere will be mixed with the light of the parent star, and it is foreseen that many transits have to be observed[23] before a good quality transmission spectrum of an exo-Earth can be obtained. Our empirical transmission spectrum suggests however, that retrieving the major planetary signals might be easier than model calculations suggest. Observations of about 20 to 30 1-hour transits should yield the detection of the major spectral features in the transmission spectrum of an Earth-like planet around a low-mass M-type star of the solar neighbourhood. With this goal in mind, is it a necessary exercise to determine and quantify the observational requirements to detect each specific molecule using the transmission spectrum presented here. It is also useful to compare the potential results from two exoplanet atmospheric characterization techniques: direct detection and transit spectroscopy.

Observations of the earthshine, the light reflected from the dark side of the Moon, are often obtained and studied as a proxy for direct extrasolar planet observations[9,24-27] mainly using chroronographic or interferometric techniques[28,29]. Here, we were able to carry out earthshine observations with the same instrumental setup as that during our lunar eclipse observations (see Supplementary Information for details about our data acquisition and analysis). The data are displayed in Figure 1b for comparison to the transmission spectrum

The depth of the features in the reflected spectrum of the Earth can vary depending on i) changes in the surface properties where part of the light is reflecting from, and ii) changes in the cloud coverage of the globe[25,30]. However, these variations are typically small (≤6%) and they do not modify significantly the shape of the data. In the transmission spectrum, there is a negligible contribution from light reflecting from the earth's surface, thus only the changes in cloudiness can affect the depth of the absorption features, and we expect the amplitude of the spectral variability to be at most of the same order than in the reflected spectrum.

In Table 1, the major molecular features in the transmission and reflected spectra, together with their equivalent widths measured over the original data with a resolution of 0.0013-0.0024 ⬚m, are listed. In the transmission spectrum, even at very low signal-to-noise ratios the major atmospheric components remain marginally detectable, but not in the reflected spectrum (in the Supplementary Information a detailed analysis of confidence of detection of atmospheric features at different signal-to-noise levels is given). Thus, the transmission spectrum can provide much more information about the atmospheric composition of a rocky planet than the reflected spectrum.

Supplementary Information is linked to the online version of the paper at www.nature.com/nature

Acknowledgements We are grateful to Drs. Franck Grundahl and Johan Fynbo for kindly allowing us access of their awarded time at the Nordical Optical Telescope, thus making this work possible. We are also grateful to Drs.V. J. S. Béjar, E. Guinan, S. Seager, B. Portmann, A. Garcia-Muñoz, and Y. Pavlenko for useful discussions. Support for this project has been provided by the Spanish Ministry of Science via the Ramon y Cajal fellowship for E.P. and project AYA2007-67458. Based on observations made with the WHT (operated by the Isaac Newton Group) and the NOT (operated by Denmark, Finland, Iceland,

Norway, and Sweden), both on the island of La Palma in the Spanish Observatorio del Roque de los

Muchachos of the Instituto de Astrofisica de Canarias.

Table 1: Pseudo-equivalent widths of the major molecular absorption features in the transmission and reflection spectra of the Earth. Pseudo-equivalent widths (given in Angstroms) are measured with respect to the pseudo-continuum that is absorbed by much wider features, like Rayleigh scattering or extended water bands. The second column is indicative of the wavelength range over which the various features are integrated. In all cases, except for a few features in the optical, the pseudo-equivalent widths are larger in the transmission spectrum than in the reflectance data. The indices -a to -f in the first column are used to compare with the filter sets defined in Table S1 in the on-line Supplementary Information.

| Atmospheric species | Wavelength interval ($\mu$m) | Equivalent width Transmission (A) | Equivalent width Reflection (A) |
|---|---|---|---|
| $O_2$-a | $0.6858 - 0.6952$ | $17.0 \pm 0.4$ | $13.5 \pm 1.0$ |
| $O_2$-b | $0.7570 - 0.7706$ | $55.0 \pm 0.3$ | $47.8 \pm 1.0$ |
| $H_2O$-a | $0.7133 - 0.7342$ | $12.2 \pm 1.0$ | $34.0 \pm 1.0$ |
| $H_2O$-b | $0.8057 - 0.8400$ | $21.2 \pm 1.0$ | $50.8 \pm 5.0$ |
| $H_2O$-c | $0.8884 - 0.9966$ | $331.1 \pm 9.0$ | $205.3 \pm 6.0$ |
| $H_2O$-d | $1.0870 - 1.1755$ | $381.0 \pm 9.0$ | $160.71 \pm 10.0$ |
| $H_2O$-e | $1.3000 - 1.5212$ | $1300.0 \pm 30.0$ | $623.4 \pm 30.0$ |
| $H_2O$-f | $1.7586 - 1.9824$ | $1111.0 \pm 30.0$ | $350.0 \pm 80.0$ |
| $O_2 \bullet O_2$-a | $0.5556 - 0.6200$ | $71.1 \pm 4.0$ | $\leq 50.0$ |
| $O_2 \bullet O_2$-b | $1.0295 - 1.0872$ | $69.0 \pm 2.0$ | $< 5.0$ |
| $O_2 \, O_2 \bullet O_2 \, O_2 \bullet N_2$-ab | $1.2347 - 1.2853$ | $178.9 \pm 9.0$ | $24.2 \pm 3.0$ |
| $CO_2$-a | $1.5629 - 1.5908$ | $52.6 \pm 1.0$ | $10.0 \pm 3.0$ |
| $CO_2$-b | $1.5908 - 1.6214$ | $54.8 \pm 1.0$ | $12.0 \pm 3.0$ |
| $CO_2$-c | $1.9906 - 2.0373$ | $150.6 \pm 1.0$ | $23.0 \pm 5.0$ |
| $CO_2$-d | $2.0373 - 2.0789$ | $127.6 \pm 1.0$ | $<15.0$ |
| $CH_4$-a | $1.6214 - 1.6750$ | $58.0 \pm 1.0$ | $<15.0$ |
| $CH_4$-bcd | $2.1390 - 2.4000$ | $>790$ | $>280.0$ |

Figure 1: The Earth's visible and near-infrared transmission and reflection spectra. The Earth's transmission spectrum is a proxy for Earth observations during a primary transit as seen beyond the Solar System, while the reflectance spectrum is a proxy for the observations of Earth as an exoplanet by direct observation after removal of the Sun's spectral features. Panel a: The transmission spectrum, with some of the major atmospheric constituents marked. The spectrum has a resolution of 0.00068 ᐧm in the optical (R~960) and 0.0013-0.0024 ᐧm in the near-infrared (R~920) A detailed atlas of the transmission spectrum with identifications of the main atomic and molecular absorption features is shown in Figure S3. Panel b: A comparison between the Earth's transmission (black) and reflexion (blue) spectra. Both spectra have been degraded to a spectral resolution of 0.02 ᐧm and normalized at the same flux value at around 1.2 ᐧm. It is readily seen from the figure that the reflected spectrum shows increased Rayleigh reflectance in the blue. It is also noticeable how most of the molecular spectral bands are weaker, and some inexistent, in the reflected spectrum. In both panels the noise (rms) of the spectra, which takes into account the corrections for the strongest local telluric features, are plotted point-per-point along the spectra (grey color) although, for most spectral regions, the size of the error bars is comparable to the width of the line. The quality of the transmission data is measured in terms of signal-to-noise ratio, which goes from ~100 (at the deepest absorptions features and the blue optical wavelengths) up to ~400 (at the largest values of the relative fluxes).

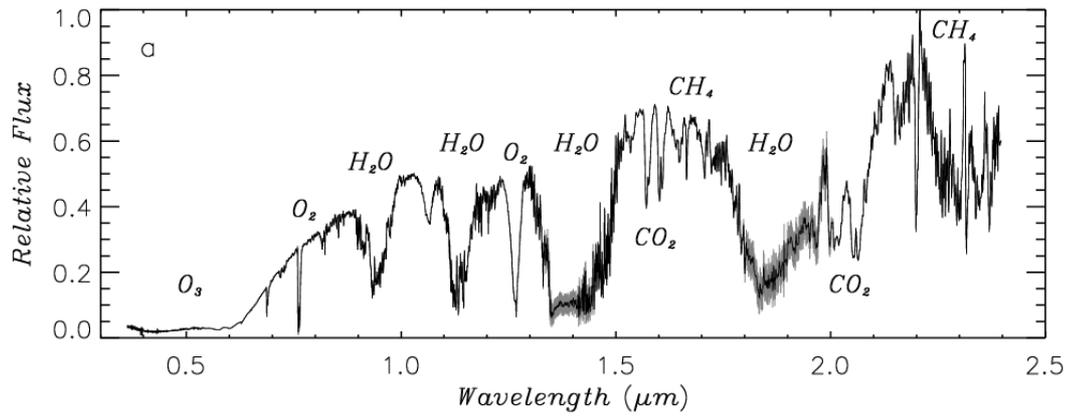

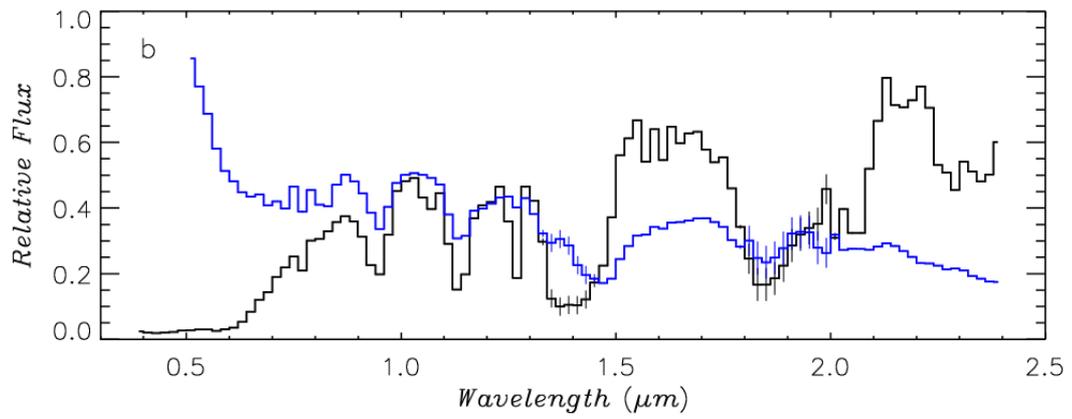

# Supplementary Information:

**Observing Methods**

Observations of the lunar eclipse of 2008 Aug 16[th] were taken simultaneously with the near-infrared LIRIS and optical ALFOSC spectrographs attached to the William Herschel Telescope (WHT) and the Nordical Optical Telescope (NOT), respectively, both at the observatory of El Roque de los Muchachos in La Palma Island.

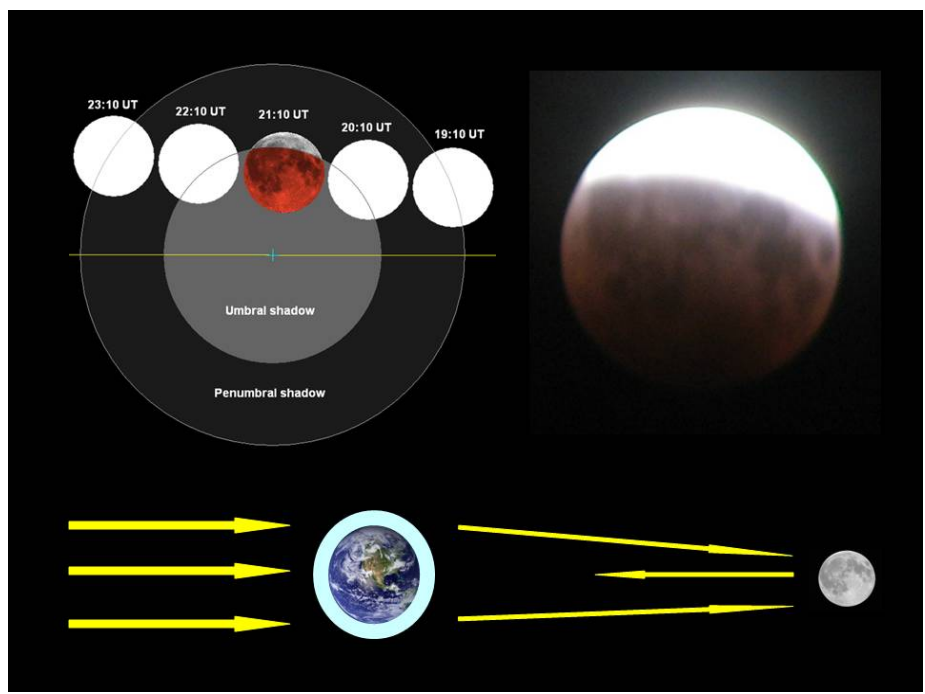

Figure S1: Details on the partial lunar eclipse on August 16[th] 2008. Top left: The temporal evolution of the eclipse (from NASA eclipse page at http://eclipse.gsfc.nasa.gov). Top right: An image of the Moon taken at 21:10 UT, the time of Greatest eclipse. Bottom: A not-to-scale cartoon illustrating the path of the sunlight through the Earth's atmosphere before reaching the Moon during the eclipse, and back to Earth. Considering the Earth's radius and the mean Sun-Earth and Earth-Moon distances, in order to reach the umbral shadow centre, the sunlight transmitted trough the Earth's atmosphere must be refracted by an angle smaller than 2°. During our observations the refraction angle was always smaller than 0.5°, practically a plane-parallel geometry, as in the case of an extrasolar planetary transit.

Observations started a few minutes before the eclipse deepest point and continued until a few hours after the end. Figure S1 depicts the sketch of the eclipse. During the time when parts of the Moon where in the umbra, the rest of the Moon's surface was located inside the penumbral region. During this time interval, the observing procedure was to alternatively observe two (umbra and penumbra) regions located near the North and South lunar poles. Past 22:44 UT, when the umbra was not visible anymore, we continued the alternate observations between two penumbra / full Moon regions with LIRIS and only full Moon regions with ALFOSC. The instrumental setups were as follows: with ALFOSC/NOT we used grism #4, the blocking order filter WG345, and a slit width of 0.4", providing a spectral resolution of 6.8 Å (R = 960 at 0.65 $\mu$m) and a wavelength coverage of 0.36-0.95 $\mu$m; with LIRIS/WHT we used grisms ZJ and HK covering 0.9-1.5 $\mu$m and 1.4-2.4 $\mu$m, respectively. We used a slit width of 0.65" giving a spectral resolution of 13 and 24 Å for each grism (R ~ 920 at 1.2 and 2.2 $\mu$m). Typical exposure times were 120 s during the eclipse and 0.2-0.5 s out of eclipse in the optical. Several (3 to 5) individual exposures were collected in the umbra, penumbra, and bright lunar regions. In the near-infrared wavelengths, we obtained several tens of images with exposure times of 40 and 15 s during the umbra observations with the ZJ and HK grisms, respectively. In order to observe the penumbra and full Moon regions within the linearity regime of the near-infrared detector of LIRIS, we used three neutral density filters to minimize the signal of the bright Moon by 2, 3 and 5 magnitudes. We also collected tens of spectra during the penumbra and bright Moon (i.e., out of eclipse). Optical and near-infrared observations were performed at airmass between 1.4 and 2.9.

The near-infrared Earth's reflected spectrum was also taken with the WHT and the same instrumental setup on June 28th 2008. The observing procedure was to take alternate observations of the dark and bright side of the Moon to obtain the Earth's reflectance as a function of wavelength. The ratio spectra between the dark and bright side of the Moon is the Earth's wavelength-dependent reflectance. As in the case of the eclipse observations, the spectral features of the Sun, the moon, and the local atmospheric airmass, are cancelled in the ratio spectra. The final reflection spectrum has S/N levels similar to those of the transmission spectrum. The detailed observational procedure and data reduction methods for the earthshine observations are described elsewhere[9,26,31,32]. On June 28th, we were not able to cover the optical spectral range between 0.5 and 0.9 $\mu$m, and to make a complete comparison between the reflected and transmitted spectra, we have included a visible earthshine spectrum[9] obtained at the same lunar phase with the 60" Telescope at Palomar observatory on Nov 19th 2003.

## Data Analysis Methods

The final set of raw spectra was reduced using standard astronomical procedures typical of optical and near-infrared wavelengths, such as dividing by the flatfield exposures and subtracting the background sky contribution. All these steps were done within the IRAF environment. Two dimension spectra were optimally extracted using the TWODSPEC package within IRAF. The number of counts of each registered individual spectrum in the umbra ranges from ~3,000 to ~8,000 counts per pixel from the blue to the red optical wavelengths, and around 15,000 counts per pixel in the near-infrared. We note that the flux at the deepest water vapour absorptions in the LIRIS data drops to ≤1,000 counts per pixel. Wavelength calibration was performed using observations of known He+Ne (optical) and Ar+Xe (near-infrared) emission lines obtained before and after the eclipse. Individual spectra of each umbra, penumbra and bright Moon regions were combined in groups of 3 to 5 to increase the signal-to-noise ratio of the data. Figure S2 shows three of these spectra (Z-J wavelengths) for the same lunar region, illustrating the marked change of some absorption features. For example, note that the feature at 1.06 μm is observable only in the umbra spectrum, and that the signature at 1.26 μm becomes significantly weaker during the bright Moon (out of eclipse) observations.

The final transmission spectrum of the Earth, depicted in Figures 1a and S3, was calculated by computing the ratio of the umbra/penumbra regions taken at the same averaged airmass in order to minimize the local atmosphere's telluric line variations. This ratio measurement effectively cancels out the atomic spectral features of the solar spectrum, the spectral albedo of the lunar surface, and local telluric absorption lines produced by the local airmass on top of the telescope. We estimate that all these corrections have been performed at a level better than 20% (local telluric lines) and 0.5% (cancellation of the solar light and the lunar albedo). For the strongest (sometimes "saturated") local telluric features (e.g., at 0.76, 1.35, and 1.83 μm), the corrections can be worse by a factor or 2.

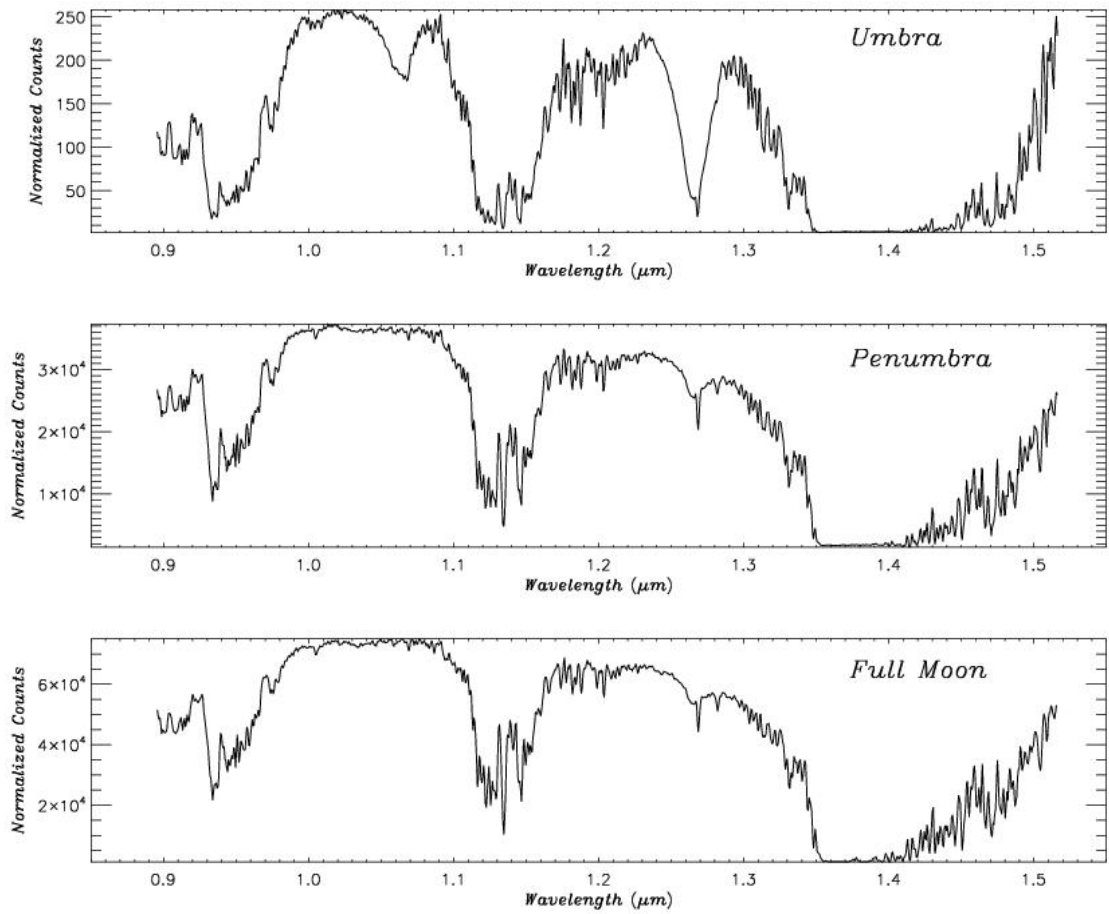

Figure S2: The observed Z-J band raw spectra of the Moon, at the same lunar location, when the region was located in the umbra, in the penumbra, and out of the eclipse (bright Moon), from top to bottom respectively.

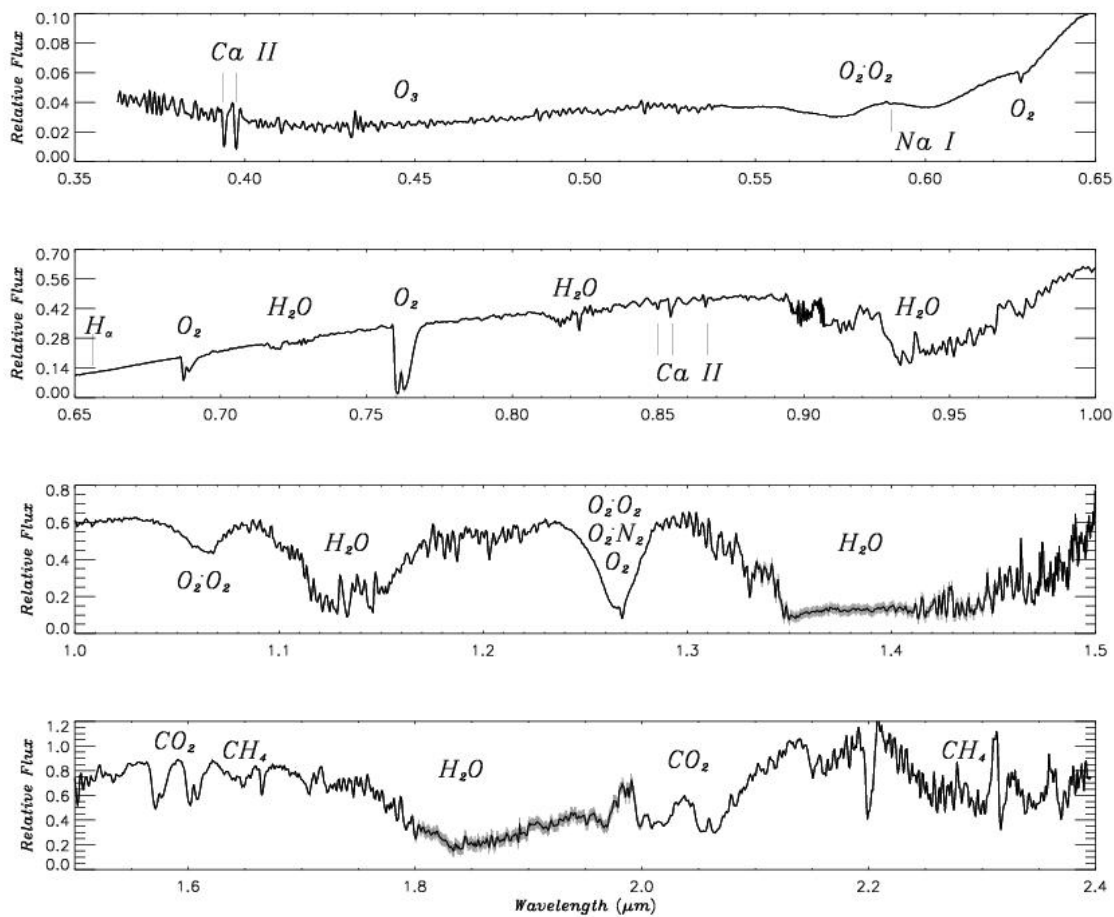

Figure S3: The Earth´s transmission spectrum from 0.36 to 2.40 ⬚m. The major atmospheric features of the spectrum are marked. On the second panel from the top, the location of the H

completely removed in the transmission spectrum. The signal-to-noise ratio (S/N) of the data as measured at the pseudo-continuum fluxes are as follows: 100-200 for the top panel, 250-400 for the second panel, and 200-400 for the third and fourth panels. At the bottom of the deepest absorption features (e.g., water vapour in the near-infrared and ozone bands), the values of S/N decrease by a factor of 2. The strong absorption features at 1.06 and 1.26 ⬚m, bands are produced by molecular oxygen ($O_2$), and oxygen collision complexes, including collisions between $O_2$ and $N_2$. The broad absorption feature of the $O_3$ Chappuis band is also clearly observable between 0.375 and 0.650 ⬚m. The relatively weak absorption lines corresponding to singly ionized calcium atoms (Ca II) at 0.3934, 0.3968 ⬚m (H and K lines), 0.8498, 0.8542, and 0.8662 ⬚m (the near-infrared triplet), originate in the uppermost layers of the Earth's atmosphere.

However, this ratio is still affected by the fact that the penumbral region has also a small contribution of refracted Earth light. This refracted light from the edge of the Earth's atmosphere is dominated by Rayleigh scattering. Thus, a further wavelength-dependent correction is needed before we obtain the true transmission spectrum. To correct our spectrum we calculated the brightness ratio of the South Pole lunar region when in the penumbra and in the full Moon (out of the eclipse). This ratio spectrum is essentially flat except for residuals of the local telluric lines (spectra were taken at different times) and a smoothly rising continuum shape, which corresponds to the characteristic wavelength dependence of $\lambda^4$ of the Rayleigh scattering, plus some remaining telluric absorption features due to the fact that the two spectra are taken at different local airmasses (different times). We adjusted a fourth order polynomial fit to this continuum, and we multiplied the umbra/penumbra spectral ratio by it, to obtain our final Earth transmission spectrum. Selecting different spatial portions of the Moon spectra does not change either the umbra/penumbra or the penumbra/full Moon ratio spectra beyond the uncertainties, either in the continuum or in the depth of the spectral bands. The quality of the final transmission spectrum is measured in terms of signal-to-noise ratio at the pseudo-continuum fluxes as follows:  100-200 in the spectral range 0.36-0.65 $\mu$m, 250-400 (0.65-1.0 $\mu$m), and 200-400 (1.0-2.4 $\mu$m). (See also caption of Figure S3).

**Main molecular and atomic features in the Earth's transmission spectrum**

In Figure S3, the major absorption features of the Earth's transmission spectrum are marked. In principle, the spectral shape of the Rayleigh scattering could be used as an indicator of the most abundant atmospheric molecular species, however, in practice the identification and quantification of the specific gases is not unequivocal[10], and can be further masked by the presence of clouds and/or aerosols. Other notorious, broad features in the Earth's transmission spectrum are seen in absorption and we have identified them as due to molecular oxygen, water vapour, methane, ozone, and carbon dioxide; these gases are small constituents of the Earth atmosphere despite the fact they all serve as biomarkers. In this sense, molecular nitrogen, $N_2$, which is actually the major gas component (78.08 % in volume) is known to lack marked electronic transitions from the ultraviolet to the near-infrared wavelengths[10], and this is certainly the case for the Earth's reflected spectra[24,25]. However, we find that observed in transmission, the Earth presents strong absorption features at 1.06 and 1.26 $\mu$m, bands likely produced by molecular oxygen ($O_2$), and oxygen collision complexes, including collisions between $O_2$ and $N_2$.

Oxygen collision complexes are van der Waals molecules, also know as dimers[16]. These loosely bounds species, held together by weak inter-molecular attractions rather than by chemical bonds, are present in all gases and have been observed experimentally in a variety of systems[17]. In the natural atmosphere, certain dimer species have mixing ratios of the same order of magnitude as minor atmospheric constituents such as carbon dioxide, ozone, water vapour or methane[16]. However, only the oxygen dimer $O_2 \bullet O_2$ and the water dimer $H_2O \bullet H_2O$ have been measured directly with long atmospheric absorption paths[16]. Laboratory work[18] shows that collision pair complexes of oxygen (which have important climate implications) are not restricted to $O_2 \bullet O_2$ and that absorption by $O_2 \bullet N_2$ is also important. Experimental data rule out the possibility that $O_2 \bullet CO_2$ or $O_2 \bullet Ar$ could significantly contribute to atmospheric absorption[18].

In the Earth's transmission spectrum, the oxygen band at 1.26 m is produced by $O$, $O_2 \bullet O_2$ and $O_2 \bullet N_2$ collision complexes, and can be used in combination with other molecular oxygen bands at 0.69, 0.76 and 1.06 m to derive an averaged atmospheric column density of $N_2$. The strength of these bands in the transmission spectrum implies that atmospheric dimers may become a major subject of study for the interpretation of rocky exoplanets transmission spectra and their atmospheric characterization. Thus, there is a clear need for further experimental developments of collision complexes molecular database[18]. Repeated lunar eclipse observations, combined with modelling efforts, can provide a tool to understand the fundamental properties of these bands.

In Figure 1b the atmospheric spectral bands of $O_3$, $O_2$, $H_2O$, $CO_2$, and $CH_4$ are readily distinguishable, even if the transmission spectrum is degraded to a spectral resolution of 0.02 m (resolving power of 50 at 1 m) as in Figure 1b. These results contrast with previous model efforts of Earth-like transit atmosphere characterizations where these spectral signatures were expected to be thinner and weaker[10]. In our data, the contrast ratio of these features in the integrated Sun+Earth spectra ranges from $10^{-5}$ to $10^{-6}$, which clearly differs with respect from the Sun-Earth flux contrast ratio of $10^{-10}$-$10^{-11}$ expected for Earth's reflected light.

Trace amounts of $N_2O$, $OClO$ and $NO_2$ (a gas mainly produced by human activities) might be present in our data, but their detection will need future detailed modelling efforts to fit the observations. They would not be detectable in an exoplanet's atmosphere if their abundance is similar as in the Earth. Nevertheless, the study of these species, which are not homogeneously distributed in altitude, will be relevant to discern how deep into the Earth´s atmosphere we are exploring. In particular, knowing if we are sampling the atmospheric features under the cloud layer is very relevant to the discussion on how feasible is the detection/characterization of super-earth atmospheres with transits[11]. Over the whole atmospheric ring that we are sampling in our observations, it is likely that at least some areas of the Earth will be cloud-free or covered by thin cirrus clouds. In principle, a contribution of light transmitted through atmospheric paths close to the Earth´s surface could be present in the spectra.

The presence of the Earth's ionosphere is also revealed in our transmission spectrum through the detection of relatively weak absorption lines corresponding to singly ionized calcium atoms (Ca II) at 0.3934, 0.3968 μm (H and K lines), 0.8498, 0.8542, and 0.8662 μm (the near-infrared triplet), which originate in the upper layers of the Earth's atmosphere. Calcium is the fifth most abundant element in the Earth's crust and the sixth most abundant one in Earth, including both crust and atmosphere. These lines as well as other solar atomic transition lines, such as the strong Hα μm, are present in all our raw spectra, but the great majority of them cancel out completely when the ratio spectrum (umbra/penumbra) is taken (see Figure S4). Thus, we can safely conclude that the atomic transitions identified in our transmission spectrum do come from the Earth's atmosphere. It is possible that other, more abundant[19] ionospheric species, such as the singly ionized magnesium (Mg II), can be detectable at shorter wavelengths that are not covered by our data. We integrated across the Ca II lines to get their equivalent widths (or strengths): 7.0 Å (Ca II H), 8.2 Å (Ca II K), 0.7 Å (0.8498 μm), 2.5 Å (0.8542 μm), and 0.80 Å (0.8662 μm), with an error estimated at 5%.

The atomic transition lines become much diluted when the transmission spectrum is degraded to low spectral resolution. However, the neutral sodium (Na I) resonance doublet at 0.5890 and 0.5896 μm has already been measured in the warm atmosphere of the transiting giant gas planet HD 209458b[3], with a photometric signature of $2.3 \times 10^{-4}$. We note that only the core of this doublet, which remains unresolved in our data, is seen in the Earth's transmission spectrum because it is embedded in the strong, broad flux absorption due to ozone, oxygen pair complexes, and the scattering by air molecules and particles. For an Earth-like planet, Ca II detection would be possible if contrast ratios of $10^{-5}$–$10^{-6}$ can be achieved.

On Earth, the mean total column density of the ionized metals is $4.4 \pm 1.2 \times 10^{9}$ cm$^{-2}$ in periods without special meteor shower activity, but increases by 1 order of magnitude during meteor showers[20]. Thus, the detection of a strong ionized metal layer in an exoplanet could be indicative of heavy meteor bombardment or strong volcanic activity. For transiting planets, it is relatively straightforward to determine a mean density[33] and to infer the presence of a heavy core. This information, combined with the characterization of the ionosphere, could also be important in quantifying the presence and nature of a planetary magnetic field.

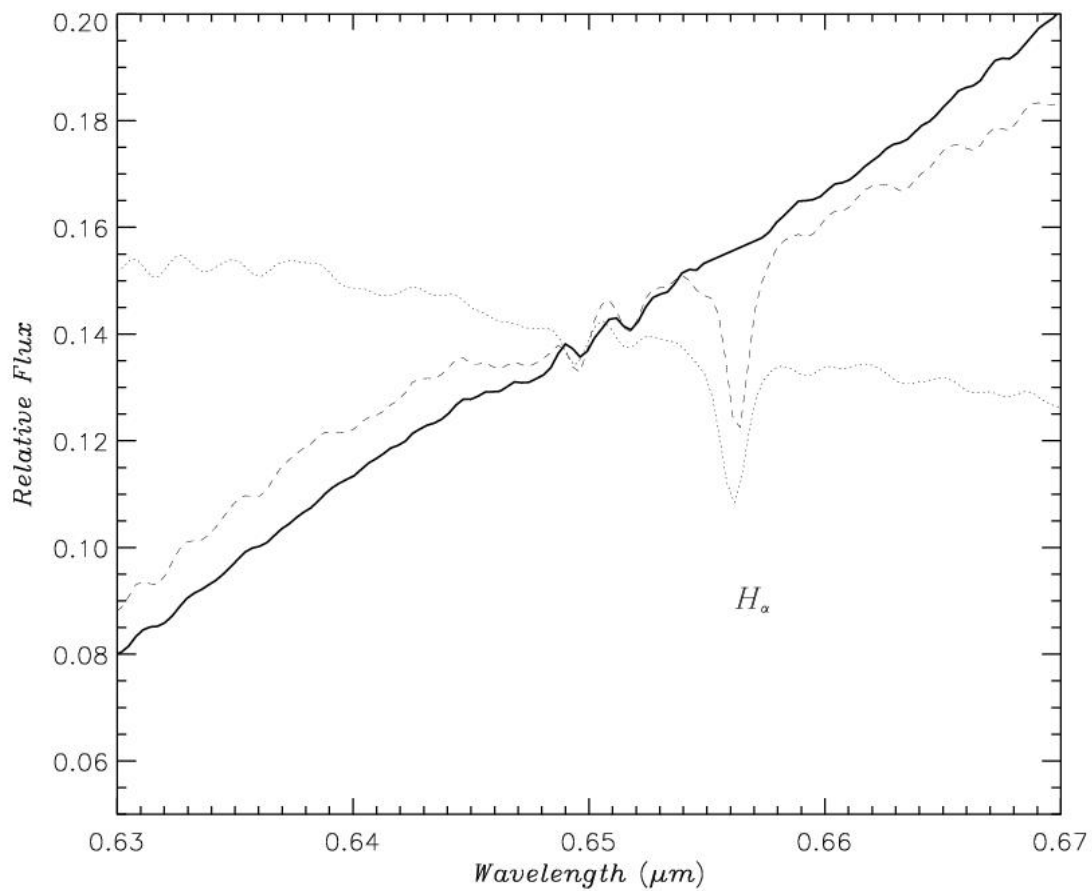

Figure S4: A detail of the umbra spectrum (broke line), the brigh Moon spectrum (dotted line) and their ratio spectrum (umbra/bright) around the hydrogen alpha ($H_t$) line (0.6568 □m). The $H_m$ solar line is a great example to illustrate the high S/N of our observations. In this figure, one can see that the $H_o$ is present in the raw spectra of the umbra and the bright Moon, but not in the final transmission spectra. The S/N ratio in all the spectra is so large that the ratio umbra/bright completely removes any contribution from the solar spectrum and the local telluric atmosphere.

**Atmospheric features detectability**

Because of the faintness of rocky planets compared to their parent stars, future observations will necessarily be photon starved, and the spectral resolution of the observations will likely be small in order to gain sufficient signal-to-noise (S/N) ratio. Thus, to aid the detectability of atmospheric molecules, we have defined a series of passbands over which the spectra can be integrated. The averaged flux is then calculated for each passband. Some passbands are used to determine the signal of the spectrum's pseudo-continuum and some are used to determine the signal of the atomic/molecular signatures. All the pass bands are 0.01 $\mu$m (100 Å) wide (see Table S1 for detailed wavelength coverage of the passbands related to molecular features), which is a relatively narrow interval especially in the infrared. In practice, broader band filters could be designed for some of the broader molecular absorptions. The passbands in Table S1 are designed to maximize the retrieval of atmospheric composition information from the Earth's transmission spectrum, and will be valid for the analysis of the transmission spectra of any extrasolar rocky planet with an atmosphere, including super-Earths.

We have applied our filters to our original spectra (transmitted and reflected) but also to two other low S/N ratio artificial spectra generated by adding different levels of random noise to the observed data. An arbitrary parameter, $\eta$, is defined to quantify the detectability of the various features according to the following dimensionless expression: $\eta = (X_c - X_m) / (sqrt(2) r_c)$, where $X_c$ and $X_m$ are the mean value of the relative fluxes in the continuum and molecular filter intervals and $r_c$ is the mean standard deviation of the flux continuum. Values of $\eta$ above 1.5 implies reliable detection; the larger the $\eta$ value, the higher confidence in the detection of the measured feature. Table S1 provides the $\eta$ values for various molecular features measured over the original spectra (S/N ratio in the range 100-400) and data whose quality has been degraded to low S/N ratios. In most cases only the closest continuum filter is used to calculate the detectability of a molecular band, but in the wavelength range between 0.6 and 1 $\mu$m, where the pseudo-continuum has a strong gradient, two continuum filters, one immediately previous and one immediately after the absorption feature, are used to estimate the detectability. From Table S1, it becomes evident how in the transmission spectrum, even at S/N=5 (i.e., poor quality data), $H_2O$, $CO_2$, $CH_4$, $O_2 \bullet O_2$ and $O_2 \bullet N_2$ remain detectable with a $\eta \geq 1.5$. In comparison, at the same S/N ratio, no molecular features can be detected with sufficient confidence in the reflected spectrum.

Table S1: The (air) wavelength range for a set of optimal visible and near-infrared filters, designed to retrieve the maximum information from the Earth's spectrum. Also given is the dimensionless detectability parameter ⬚ (equation defined in the text of the on-line material) for each molecular band observed in the transmission and reflected spectra. The larger the ⬚ value, the higher confidence in the detection of the measured feature. For each case, three different S/N ratios (our original spectrum, S/N of 10 and 5) have been considered by adding artificial noise to our data. In the table, ⬚ is only given when its value is equal to or larger than 1.5. The pseudo-continuum spectral bands of reference are 0.01 ⬚m (100 Å) wide with initial wavelengths at 0.540, 0.640, 0.660, 0.700, 0.735, 0.780, 0.880, 1.025, 1.080, 1.230, 1.290, 1.550, 1.675, 2.130, and 2.210 ⬚m.

| Atmospheric species | Filter wavelength range (⬚m) | Transmission spectrum | | | Reflection spectrum | | |
|---|---|---|---|---|---|---|---|
| | | S/N > 100 | S/N=10 | S/N=5 | S/N > 100 | S/N=10 | S/N=5 |
| $H_2O$-a | 0.685 – 0.695 | 2.4 | - | - | 2.9 | - | - |
| $H_2O$-b | 0.810 – 0.820 | 6.0 | - | - | 9.2 | - | - |
| $H_2O$-c | 0.930 – 0.940 | 28.1 | 2.1 | - | 8.6 | 1.6 | - |
| $H_2O$-d | 1.120 – 1.130 | 44.0 | 3.1 | 1.5 | 10.0 | - | - |
| $H_2O$-e | 1.350 – 1.360 | 53.5 | 3.8 | 2.0 | 6.5 | - | - |
| $H_2O$-f | 1.830 – 1.840 | 69.4 | 4.7 | 2.3 | 8.2 | 1.5 | - |
| $O_2 \bullet O_2$-a | 0.570 – 0.580 | 12.0 | - | - | - | - | - |
| $O_2$-a | 0.758 – 0.768 | 26.1 | 1.5 | - | 8.0 | 2.5 | - |
| $O_2 \bullet O_2$-b | 1.060 – 1.070 | 23.0 | - | - | 2.0 | - | - |
| $O_2\ O_2 \bullet O_2\ O_2 \bullet N_2$-a | 1.254 – 1.264 | 29.0 | 2.4 | - | 2.6 | - | - |
| $O_2\ O_2 \bullet O_2\ O_2 \bullet N_2$-b | 1.265 – 1.275 | 35.6 | 2.8 | 1.5 | 5.8 | - | - |
| $CO_2$-a | 1.570 – 1.580 | 23.0 | 2.1 | - | - | - | - |
| $CO_2$-b | 1.600 – 1.610 | 21.8 | 1.6 | - | - | - | - |
| $CO_2$-c | 2.010 – 2.020 | 24.2 | 5.1 | 2.3 | 1.8 | - | - |
| $CO_2$-d | 2.060 – 2.070 | 26.1 | 5.3 | 2.3 | - | - | - |
| $CH_4$-a | 1.640 – 1.650 | 9.0 | - | - | 1.7 | - | - |
| $CH_4$-b | 2.230 – 2.240 | 2.7 | 2.2 | 1.6 | - | - | - |
| $CH_4$-c | 2.250 – 2.260 | 5.4 | 3.4 | 2.2 | 2.1 | - | - |
| $CH_4$-d | 2.290 – 2.300 | 6.6 | 4.2 | 2.7 | 2.7 | - | - |

Thus, the transmission spectrum can provide much more information about the atmospheric composition of a rocky planet than the reflected spectrum, and roadmaps leading to the detection of exoplanet atmospheres may be designed accordingly. Nevertheless, the direct detection of exoplanets is still worth pursuing as it can give highly valuable information, such as diurnal, seasonal and orbital variability[30], or some selected surface properties (vegetation's red edge, polarization, ocean glint, etc.), which cannot be obtained from the transmission spectrum. Furthermore, transmission spectra can only be obtained under favourable geometric circumstances, which are not very frequent.

**Supplementary Bibliography**